\documentclass[showpacs,preprint]{revtex4}
\usepackage{hyperref}
\usepackage{amsfonts}
\usepackage{amsmath}
\usepackage{amssymb}
\usepackage{graphicx}

\begin{document}

\title{First and second laws of modified gravity theories on dynamical
holographic screens}
\author{Shao-Feng Wu$^{1,2}$\footnote{%
Corresponding author. Email: sfwu@shu.edu.cn; Phone: +86-021-66136202.},
Xian-Hui Ge$^{1,2}$, Peng-Ming Zhang$^{3,4}$, Guo-Hong Yang$^{1,2}$}
\affiliation{$^{1}$Department of physics, Shanghai University, Shanghai, 200444, P. R.
China}
\affiliation{$^{2}$The Shanghai Key Lab of Astrophysics, Shanghai, 200234, P. R. China}
\affiliation{$^{3}$Center of Theoretical Nuclear Physics, National Laboratory of Heavy
Ion Accelerator, Lanzhou 730000, P. R. China}
\affiliation{$^{4}$Institute of Modern Physics, Lanzhou, 730000, P. R. China}

\begin{abstract}
We show that field equations of some typical modified gravity theories can
be recast as thermodynamic first law on dynamical holographic screens. The
thermodynamic second law are also presented.
\end{abstract}

\maketitle

\section{Introduction}

Since the discovery of the black hole entropy and the analogue between the
laws of black hole mechanics and thermodynamics \cite{Bekenstein,Bardeen},
there are increased interest on the thermodynamic feature of gravity \cite%
{PadRPP}. It is generally believed that the puzzling feature should be
clarified in quantum gravity. For instance, it is often claimed that one of
the greatest triumphs of string theory is its ability to recover the
area-entropy relation from the counting of string microstates. In addition,
gravitational thermodynamics has stimulated some intriguing ideas about the
spacetime or gravity, such as the holographic principle \cite{Hooft,Susskind}%
, its possible realization, namely, the AdS/CFT correspondence, and the
growing consensus that gravity might not be a fundamental interaction but
rather an emergent large scale/numbers phenomenon.

In general, the mentioned black holes, or the gravitational thermodynamic
systems, have been defined theoretically in terms of event horizons, usually
the past causal boundary of future null infinity, with temperature
proportional to surface gravity and entropy measured by its horizon area.
The event horizon is treated as a real thermodynamic system after the
discovery of quantum Hawking radiation \cite{Hawking} with temperature
proportional to surface gravity on the horizon. To understand Hawking
radiation, Unruh \cite{Unruh} considered the Rindler horizon, which is
witnessed by an accelerating observer on Minkowski spacetime, who feels the
traditional vacuum state as a thermal spectrum of particles with temperature
proportional to the acceleration. On the local Rindler horizon in curved
spacetime, Jacobson disclosed that Einstein's equation can emerge as an
equation of state from the Clausius relation assuming the area-entropy
relation and Unruh temperature on the horizon \cite{Jacobson}. Respecting
that the preferred time direction, namely the Killing vector in stationary
spacetimes, should be replaced with the Kodama vector \cite{Kodama} in
dynamical spacetimes, Hayward \cite{Hayward1,Hayward0} constructed the
thermodynamic laws on the trapping horizon which is a spherically dynamical
surface with null expansion. Interestingly, in term of the tunneling
approach proposed by Parikh and Wilczek \cite{Wilczek}, it has been shown
the Hawking radiation with temperature proportional to surface gravity on
trapping horizon \cite{CriscienzoPLB,CriscienzoCQG}. Hence it seems
reasonable that various horizons behave as the ordinary thermodynamic
systems. Moreover, the rationality is also supported by the fact that these
horizon thermodynamics can be extended beyond Einstein gravity \cite%
{Eling,Akbar,Cao,Gong,Elizalde,Cao07,Wu08,Cao3,Zhu,Wu10,Cai10}.

However, a natural question arises: can the gravity open up its
thermodynamic feature only on the horizon? Concretely, are there any
thermodynamic parameters and laws off the horizon? In \cite{Makela} a
spacetime foam model of the Schwarzschild horizon was constructed to recover
the Bekenstein-Hawking entropy, but the model implies that every spatial
two-surface of spacetime has the area-entropy relation. In \cite{Fursaev},
the entanglement entropy of fundamental degrees of freedom spatially
separated by a surface has the Bekenstein-Hawking form. In static spacetimes
the surface can be defined as a minimal hypersurface of a least volume
separating the system in a constant time slice. Recently, Verlinde \cite%
{Verlinde} claimed that gravity (both Einstein and Newton gravity) may be
explained as entropic force and space may be emergent. This remarkable work
has attracted great interest in various aspect of physics \cite{EntropyForce}%
. One of the interesting elements in this work is to associate temperature
and entropy to the holographic screens. The screens are located at
equipotential surfaces, which are not necessary the black hole horizon, but
can be some general surfaces even in the flat nonrelativistic space.

Motivated by the concept of holographic screen, the minimal surfaces have
been treated as a sort of holographic screens and the variation of its area
is proved to be proportional to the change of the distance between the test
particle and the surface \cite{Fursaev1}, which is the one of the hypothesis
given by Verlinde. Moreover, certain timelike screen is also proposed to
repeat the local Rindler horizon thermodynamics \cite{Piazza}. Recently, it
was shown that both field equations of Einstein gravity and Lovelock gravity
can be recast as the first law of thermodynamics on the general holographic
screen in the static spacetime with spherical symmetry \cite{Tian}. All
these results imply that the negative answer of the aforementioned question
is possible.

In this paper, we will develop the gravitational thermodynamics on dynamical
holographic screens. We will construct the first and second laws,
respectively for three typical modified gravity theories, i.e. the
Gauss-Bonnet gravity, $f(R)$ gravity and scalar-tensor gravity.

\section{Thermodynamical parameters on dynamical holographic screens}

Let us introduce the holographic screens in dynamical spacetime and
thermodynamical parameters we are interested in. Suppose the $n$-dimensional
spacetime ($M_{n}$, $g_{\mu \nu }$) to be a warped product of an ($n-2$%
)-dimensional spherical symmetry space ($K_{n-2}$, $\gamma _{ij}$) and a
two-dimensional orbit spacetime ($M_{2}$, $h_{ab}$) under the isometries of (%
$K_{n-2}$, $\gamma _{ij}$). Namely, the line element is given by%
\begin{equation*}
ds^{2}=h_{ab}dx^{a}dx^{b}+r^{2}(x)\gamma _{ij}dy^{i}dy^{j},
\end{equation*}%
where $r$ is the areal radius for an ($n-2$)-sphere $K_{n-2}$. It is useful
to locally rewrite the line element in the double-null coordinates as%
\begin{equation}
ds^{2}=-2e^{-\phi (u,v)}dudv+r^{2}(u,v)d^{2}\Omega _{n-2}  \label{doublenull}
\end{equation}%
where $d^{2}\Omega _{n-2}$ denotes the line element of the ($n-2$)-sphere.
The causal structure of this spacetime is convenient to be studied using
null geodesics. The null expansions of two independent future-directed
radial null geodesics are expressed as $\theta _{+}=(n-2)r^{-1}r,_{u}$ and $%
\theta _{-}=(n-2)r^{-1}r,_{v}$. An ($n-2$)-dimensional sphere is called as
marginal if $\theta _{+}\theta _{-}=0$, trapped if $\theta _{+}\theta _{-}>0$%
, and untrapped if $\theta _{+}\theta _{-}<0$. The trapping horizons \cite%
{Hayward1} are defined as the hypersurfaces foliated by marginal surfaces. A
marginal sphere with $\theta _{+}=0$ is called future if $\theta _{-}<0$,
past if $\theta _{-}>0$, bifurcating if $\theta _{-}=0$, outer if $\partial
_{v}\theta _{+}=(n-2)r^{-1}r,_{uv}<0$, inner if $\partial _{v}\theta _{+}$ $%
>0$ and degenerate if $\partial _{v}\theta _{+}=0$. On an untrapped sphere,
fixing a spatial or null vector $\xi $ is outgoing if $\xi ^{a}\nabla
_{a}r>0 $ and ingoing if $\xi ^{a}\nabla _{a}r<0$. Equivalently, fixing the
orientation locally by $\theta _{+}>0$ and $\theta _{-}<0$, $\xi $ is
outgoing if $\xi _{u}>0$ or $\xi _{v}<0$ and ingoing if $\xi _{u}<0$ or $\xi
_{v}>0$. We also call the untrapped sphere as outer, inner and degenerate
similar to the marginal surface.

In stationary spacetimes, the Killing vecto takes essential role in the
definition of some concepts one needs to discuss the gravitational
thermodynamics, including the holographic screen, surface gravity, Wald
horizon entropy, and Komar energy. In dynamical spacetimes with sphere
symmetry, there is a preferred time direction given by the Kodama vector
\cite{Kodama}, which is a natural dynamical analogue of a stationary Killing
vector. The Kodama vector is given by%
\begin{equation}
K^{\mu }\equiv -\epsilon ^{\mu \nu }\nabla _{\nu }r=(e^{\phi }\partial
_{v}r,-e^{\phi }\partial _{u}r,0,0,\cdots ),  \label{Kodama}
\end{equation}%
where $\epsilon _{\mu \nu }=\epsilon _{ab}\left( dx^{a}\right) _{\mu }\left(
dx^{b}\right) _{\nu }$, $\epsilon _{ab}$ is a volume element of ($M_{2}$, $%
h_{ab}$). It reduces to the Killing vector $K^{\mu }=(1,0,\cdots )$ for a
static spacetime%
\begin{equation}
ds^{2}=-g(r)dt^{2}+\frac{1}{g(r)}dr^{2}+r^{2}d^{2}\Omega _{n-2}.
\label{staticds}
\end{equation}

As suggested in \cite{Cao1}, we will consider holographic screens located at
equipotential surfaces defined by%
\begin{equation}
-K^{\mu }K_{\mu }\equiv e^{2\psi }=-2e^{\phi }r,_{u}r,_{v}=\text{constant }c%
\text{.}  \label{KK}
\end{equation}%
In the static spacetime (\ref{staticds}), $\psi $ is the generalized Newton
potential in Einstein gravity. It should be stressed that this definition of
holographic screens implies that we are considering the screen as untrapped
spheres with $-K^{\mu }K_{\mu }=c>0$. One will find that it is important to
prove the second law of entropy-like function on the dynamical screens.

The dynamical surface gravity associated with the trapping horizon has been
defined directly from the Kodama vector
\begin{equation}
\kappa \equiv -\frac{1}{2}\epsilon ^{ab}\nabla _{a}K_{b}  \label{Kapa}
\end{equation}%
with $c=0$. We will define the temperature of the dynamical holographic
screens as
\begin{equation}
T=\left. \frac{\kappa }{2\pi }\right\vert _{c=const}.  \label{T}
\end{equation}%
{In the static spacetime (\ref{staticds}), this temperature will reduce to
the Verlinde-Unruh temperature \cite{Verlinde}.}

Similar to the Killing vector, the Kodama vector is also related to certain
gravitational energy $E$. One can define a current by the energy-momentum
tensor of matter $J_{E}^{\mu }=T^{\mu \nu }K_{\nu }$. If there is a
conservation law $\nabla _{\mu }J_{E}^{\mu }=0$, an associated charge can be
obtained as%
\begin{equation}
E=\int_{\Sigma }J_{E}^{\mu }d\Sigma _{\mu }  \label{E}
\end{equation}%
by integrating the locally conserved currents over some spatial volume $%
\Sigma $ with boundary. For Einstein gravity, $E$ is just the well-known
Misner-Sharp energy \cite{Misner}.

For Gauss-Bonnet energy, by using field equations%
\begin{equation}
G_{\mu \nu }+\alpha H_{\mu \nu }=8\pi T_{\mu \nu },  \label{GGB}
\end{equation}%
\begin{equation*}
H_{\mu \nu }=2(RR_{\mu \nu }-2R_{\mu \lambda }R_{\nu }^{\lambda
}-2R^{\lambda \rho }R_{\mu \lambda \nu \rho }+R_{\mu }^{\;\lambda \rho
\sigma }R_{\nu \lambda \rho \sigma })-\frac{1}{2}g_{\mu \nu }L_{GB},
\end{equation*}%
the generalized Misner-Sharp energy \cite{Maeda} has been obtained%
\begin{equation}
E=\frac{(n-2)\Omega _{n-2}r^{n-3}}{16\pi }\left[ (1+2e^{\phi }r,_{u}r,_{v})+%
\tilde{\alpha}r^{-2}(1+2e^{\phi }r,_{u}r,_{v})^{2}\right] ,  \label{EGB}
\end{equation}%
where $\Omega _{n-2}$ denotes the unit area of ($n-2$)-sphere $K_{n-2}$ and $%
\tilde{\alpha}=(n-3)(n-4)\alpha $.

For nonlinear gravity theory with $L=f(R)$ and field euqations%
\begin{equation}
f_{R}R_{\mu \nu }-\frac{1}{2}fg_{\mu \nu }-\nabla _{\mu }\nabla _{\nu
}f_{R}+g_{\mu \nu }\square f_{R}=8\pi T_{\mu \nu },  \label{GfR}
\end{equation}
although the energy current $J_{E}^{\mu }$ is not always divergence-free,
the conservative quasi-local energy have been found \cite{Cao3} for an FRW
universe and static spherically symmetric solutions with constant scalar
curvature, resepectively. For an FRW universe with the line element%
\begin{equation*}
ds^{2}=-dt^{2}+a^{2}(t)d\bar{r}^{2}+\bar{r}^{2}a^{2}(t)d^{2}\Omega _{2},
\end{equation*}%
the generalized Misner-Sharp energy is%
\begin{equation}
E=\frac{\left( \bar{r}a\right) ^{3}}{12}\left[ f+6H\dot{f}_{R}-6f_{R}(H^{2}+%
\dot{H})\right] .  \label{EFRW}
\end{equation}%
For static spherically symmetric solutions, the energy for $n=4$ is found as%
\begin{equation}
E=\frac{r}{2}\left[ f_{R}-hf_{R}+\frac{1}{6}r^{2}(f-f_{R}R)\right] ,
\label{EfR}
\end{equation}%
where $R$, $f_{R}$, and $f$ are all needed to be constants.

For the scalar-tensor gravity with Lagrangian%
\begin{equation}
L=F(\Phi )R-\frac{1}{2}\left( \nabla \Phi \right) ^{2}-V(\Phi )  \label{LF}
\end{equation}%
where $F(\Phi )$ is an arbitrary positive continuous function of the scalar
field $\Phi $ and $V(\Phi )$ is its potential, the equations of motion are%
\begin{equation}
FG_{\mu \nu }-\nabla _{\mu }\nabla _{\nu }F+g_{\mu \nu }\square F-\frac{1}{2}%
\left[ \nabla _{\mu }\Phi \nabla _{\nu }\Phi -g_{\mu \nu }(\frac{1}{2}\nabla
_{\lambda }\Phi \nabla ^{\lambda }\Phi +V)\right] =8\pi T_{\mu \nu },
\label{GST}
\end{equation}%
\begin{equation}
\square \Phi -V^{\prime }(\Phi )+F^{\prime }(\Phi )R=0.  \label{EOM}
\end{equation}%
The generalized Misner-Sharp energy in an FRW spacetime has been obtained as%
\begin{equation}
E=\frac{\left( \bar{r}a\right) ^{3}}{12}\left[ 6\left( FH^{2}+H\dot{F}%
\right) -\frac{1}{2}\dot{\Phi}^{2}-V\right] .  \label{EST}
\end{equation}

The Misner-Sharp energy satisfies the unified first law, which was
previously proposed in Einstein gravity \cite{Hayward0}, holds also for
Gauss-Bonnet gravity, $f(R)$ gravity and scalar-tensor gravity in the
mentioned cases \cite{Cao}, with the uniform%
\begin{equation}
\nabla _{a}E=A\Psi _{a}+W\nabla _{a}V,  \label{UFL}
\end{equation}%
where $A=\Omega _{n-2}r^{n-2}$ is the area of the sphere with radius $r$ and
$V$ is its volume. $W$ is called work density defined as $W=-h_{ab}T^{ab}/2$
and%
\begin{equation}
\Psi _{a}=T_{a}^{b}\partial _{b}r+W\partial _{a}r  \label{psi}
\end{equation}%
is the energy supply vector, with $T_{ab}$ being the projection of the $n$%
-dimensional energy-momentum tenor of matter in the normal direction of the $%
\left( n-2\right) $-dimensional sphere.

Besides temperature and energy, we also need to discuss another
thermodynamical parameter, the gravitational entropy. It is known that the
entropy of stationary horizon is well defined by Wald entropy \cite{Wald1},
which is a Noether charge associated with the Killing vector, but it is less
understood for the horizon entropy in a dynamical spacetime, where the
Killing vector can not be found in general. Iyer and Wald proposed that one
can approximate the metric by its boost-invariant part to "create a new
spacetime" where there is a Killing vector. However, the obtained dynamical
entropy is not invariant under field redefinition in general \cite{Wald2}.
Hayward proposed that the Wald entropy can be alternatively associated with
Kodama vector \cite{Hayward1,Hayward2}. For Einstein gravity, the dynamical
horizon entropy, which has been called as Wald-Kodama entropy, has the same
simple form as for stationary black holes. This was also justified by
evaluating the surface terms in a dual-null form of the reduced action in
two dimensions \cite{Hayward3}. Following Hayward's proposal, we have given
a general expression of Wald-Kodama entropy on the horizon in generalized
gravity theories \cite{Wu10}. Now we present that there is the similar
expression on the off-horizon screen. We briefly give the result. For a
generally covariant total Lagrangian $L_{t}$ (including the gravity $L_{g}$
and matter contributions $L_{m}$) involves no more than quadratic
derivatives of metric $g_{\mu \nu }$ and the first order derivative of some
scalar fields $\Phi _{(i)}$, one can find a Noether current generated by the
variation induced by arbitrary vector $\varsigma ^{\mu }$%
\begin{equation}
J_{S}^{\mu }=\varsigma ^{\mu }L_{t}-\Theta ^{\mu },  \label{JS}
\end{equation}%
where%
\begin{equation}
\Theta ^{\beta }=-2X_{(\mu \;\;\nu )}^{\;\;\alpha \beta }\nabla _{\alpha
}\delta g^{\mu \nu }+2\nabla _{\alpha }X_{(\mu \;\;\nu )}^{\;\;\alpha \beta
}\delta g^{\mu \nu }+\omega _{(i)}^{\beta }\delta \Phi _{(i)},
\end{equation}%
\begin{equation*}
X^{\mu \nu \lambda \rho }=\frac{\partial L_{t}}{\partial R_{\mu \nu \lambda
\rho }},\;\omega _{(i)}^{\mu }=\frac{\partial L_{t}}{\partial \nabla _{\mu
}\Phi _{(i)}},\;\delta g^{\mu \nu }=-2\nabla ^{(\nu }\varsigma ^{\mu )}.
\end{equation*}%
Furthermore, there is an antisymmetric Noether potential satisfied with $%
J_{S}^{\mu }$ $=$ $\nabla _{\nu }Q^{\mu \nu }$. The full Noether potential $%
Q^{\mu \nu }$ can be calculated in a straightforward manner for a given
action, as shown in \cite{KMaeda,Cardoso}, which is%
\begin{equation}
Q^{\mu \nu }=-2X^{\mu \nu \lambda \rho }\nabla _{\lambda }\varsigma _{\rho
}+4\varsigma _{\rho }\nabla _{\lambda }X^{\mu \nu \lambda \rho },
\label{Jab}
\end{equation}%
Integrating the Noether potential over any closed spacelike surface $B$ of
codimension $n-2$, the Noether charge is proportional to%
\begin{equation}
S=\frac{1}{8\kappa }\int_{B}Q^{\mu \nu }dB_{\mu \nu },  \label{WaldS}
\end{equation}%
where $dB_{\mu \nu }=\frac{1}{2}\epsilon _{\mu \nu }\sqrt{\gamma }d^{n-2}y$.
When $\varsigma ^{\mu }$ is a timelike Killing vector and choose $B$ as
Killing horizon, the term proportional to $\varsigma ^{\mu }$ of $Q^{\mu \nu
}$ is absent in the integral, because Killing vector vanishes on the Killing
horizon. Then the function $S$ is reduced to the Wald horizon entropy which
was used recently in \cite{Brustein}. For our aim, we will take $\varsigma
^{\mu }$ as the Kodama vector $K^{\mu }$ and $B$ as dynamical screens, so
this term must be preserved.

For Gauss-Bonnet gravity, one has%
\begin{equation}
X^{\mu \nu \lambda \rho }=g^{\mu \lbrack \lambda }g^{\left\vert \nu
\right\vert \rho ]}+2\alpha \left( g^{\mu \lbrack \lambda }g^{\left\vert \nu
\right\vert \rho ]}R+2g^{\nu \lbrack \lambda }R^{\left\vert \mu \right\vert
\rho ]}-2g^{\mu \lbrack \lambda }R^{\left\vert \nu \right\vert \rho
]}+R^{\mu \nu \lambda \rho }\right) .  \label{PGB}
\end{equation}%
For simplicity, we consider only the case with $n=5$. Using Eqs. (\ref%
{Kodama}), (\ref{Kapa}), (\ref{Jab}) and (\ref{PGB}), we can evaluate the
Noether charge (\ref{WaldS}), which is%
\begin{equation}
S=\frac{1}{4}A+3\alpha \frac{1+2\alpha e^{\phi }r,_{u}r,_{v}}{r^{2}}A=\frac{1%
}{4}A+3\alpha \frac{1-c}{r^{2}}A.  \label{SGB}
\end{equation}%
On the trapping horizon $c=0$, the entropy has the same form as the
stationary case. For $f(R)$ gravity, we have%
\begin{equation*}
X^{\mu \nu \lambda \rho }=f_{R}g^{\mu \lbrack \lambda }g^{\left\vert \nu
\right\vert \rho ]}.
\end{equation*}%
Considering a four-dimensional spacetime, we evaluate Eq. (\ref{WaldS}) as%
\begin{equation}
S=\frac{1}{4}(f_{R}-\frac{r,_{v}f_{R},_{u}+r,_{u}f_{R},_{v}}{r,_{uv}})A=%
\frac{1}{4}(f_{R}-\frac{r,_{v}f_{R},_{u}-c\frac{e^{-\phi }f_{R},_{v}}{2r,_{v}%
}}{r,_{uv}})A.  \label{SfR}
\end{equation}%
For the scalar-tensor gravity with%
\begin{equation*}
X^{\mu \nu \lambda \rho }=Fg^{\mu \lbrack \lambda }g^{\left\vert \nu
\right\vert \rho ]},
\end{equation*}%
we have obtained a similar form as Eq. (\ref{SfR})%
\begin{equation}
S=\frac{1}{4}(F-\frac{r,_{v}F,_{u}+r,_{u}F,_{v}}{r,_{uv}})A=\frac{1}{4}(F-%
\frac{r,_{v}F,_{u}-c\frac{e^{-\phi }F,_{v}}{2r,_{v}}}{r,_{uv}})A.
\label{SST}
\end{equation}%
It should be noticed that for the horizon entropy given by the
boost-invariant fields, the entropy of scalar-tensor gravity is generally
different with the entropy of $f(R)$ gravity \cite{Wald2}. But we have
argued \cite{Wu10} that our result is reasonable since the $f(R)$ gravity
can be treated as a special scalar-tensor theory by introducing the scalar
field $\phi =R$ and potential $V=\phi f^{\prime }-f$ in the Brans-Dick
theory, and choosing the Brans-Dick parameter $\omega =0$ (see \cite{Faraoni}
for a review). Moreover, one can find that even on horizons, the dynamical
entropy of $f(R)$ gravity and scalar-tensor gravity have different forms
with their stationary cases, contrary to the assumption given in many
references \cite{Eling,Akbar,Cao,Gong,Elizalde,Wu08,Cao3,Zhu}. We have shown
\cite{Wu10} that the Wald-Kodama entropy on trapping horizons preserves the
second law and can be used to construct the first law for Gauss-Bonnet
gravity on any dynamical horizons, and also for $f(R)$ gravity and
scalar-tensor gravity on slowly varying horizons, without invoking the
non-equilibrium entropy production \cite{Eling,Akbar,Cao,Cao3} and the new
mass-like function \cite{Gong,Wu08,Zhu}. These results support the
Wald-Kodama entropy as a preferred dynamical entropy expression. In the
following sections, we will further show that the Noether charge (\ref{WaldS}%
) not only has the corresponding second law but also is applicable to
construct a similar identity of first law on general holographic screens. So
we call it as the entropy on holographic screens.

\section{First law of modified gravity theories on dynamical holographic
screens}

There are two kinds of first laws (Gibbs equations) of modified gravity
theories on the trapping horizons. We have called them Hayward's identity
and Padmanabhan's identity, respectively \cite{Wu10}. We will show that both
identities can still be constructed on general holographic screens with $c$
as a constant.

\subsection{Hayward's identity of Gauss-Bonnet gravity}

Projecting the unified first law (\ref{UFL}) along a tangent vector $\xi $
of holographic screens
\begin{equation}
\xi ^{a}\nabla _{a}E=A\Psi _{a}\xi ^{a}+W\xi ^{a}\nabla _{a}V.  \label{UFL1}
\end{equation}%
We will prove Hayward's identity%
\begin{equation}
\xi ^{a}\nabla _{a}E=T\xi ^{a}\nabla _{a}S+W\xi ^{a}\nabla _{a}V,
\label{HGib}
\end{equation}%
by showing $A\Psi _{a}\xi ^{a}=T\xi ^{a}\nabla _{a}S$. The components of
tangent vector $\xi =d/d\lambda =\beta \partial _{u}-\gamma \partial _{v}$
can be determined by $\xi ^{a}\partial _{a}(K^{\mu }K_{\mu })=0$ up to a
normalization of $\xi ^{a}$, which is irrelevant for present aim. In
double-null coordinates, we write it clearly%
\begin{eqnarray}
\beta &=&cr,_{v}\phi ,_{v}+cr,_{vv}-2e^{\phi }\left( r,_{v}\right)
^{2}r,_{uv},  \notag \\
\gamma &=&cr,_{v}\phi ,_{u}+cr,_{uv}-2e^{\phi }\left( r,_{v}\right)
^{2}r,_{uu}.  \label{zeta}
\end{eqnarray}%
Using the field equation of Gauss-Bonnet gravity (\ref{GGB}), the energy
supply along the holographic screen is obtained as%
\begin{equation}
A\Psi _{a}\xi ^{a}=\frac{3\pi r,_{uv}e^{-\phi }\left[ 2(1-c)\alpha +r^{2}%
\right] }{8r,_{v}}\left[ c^{2}r,_{v}\phi ,_{v}+c^{2}r,_{vv}+2e^{\phi
}c\left( r,_{v}\right) ^{3}\phi ,_{u}-4e^{2\phi }\left( r,_{v}\right)
^{4}r,_{uu}\right]  \label{Apsi}
\end{equation}%
One can find that $\xi ^{a}\nabla _{a}S$ just equals to Eq. (\ref{Apsi}) up
to the temperature factor $T=\kappa /(2\pi )$. So we can recast the unified
first law (\ref{UFL1}) along the holographic screen as Hayward's identity (%
\ref{HGib}).

\subsection{Padmanabhan's identity of Gauss-Bonnet gravity}

We will show that there is another Gibbs equation on holographic screens
\begin{equation}
dE=TdS+WdV,  \label{PGib}
\end{equation}%
which has been called as Padmanabhan's identity. Although it is similar to
Eq. (\ref{HGib}), some differences should be clarified. The differentials $d$%
, which are different with $\xi ^{a}\nabla _{a}$, are interpreted as $%
dE=\left( dE/dr_{+}\right) dr_{+}$ etc., where $r_{+}$ refers to the radius
of holographic screen. So one must be careful that here all quantities
should be evaluated on a screen before manipulating the differentials.
Moreover, it implies that Padmanabhan's identity is only effective for the
case where $E$ and $S$ can be written as a functional of $r_{+}(u,v)$. It
also should be noticed that Padmanabhan's identity can be reduced to static
spacetime directly. However, Hayward's identity is trivial in the static
spacetime (\ref{staticds}) where $\Psi _{a}\chi ^{a}=0$ since $\chi ^{a}$
only has time component while the time component of $\Psi _{a}$ is vanishing.

The work density on the screen can be derived as%
\begin{equation}
W=e^{\phi }T_{uv}=\frac{3}{8\pi r^{3}}\left[ (1-c)r+e^{\phi
}r^{2}r,_{uv}+4(1-c)\alpha e^{\phi }r,_{uv}\right] _{r=r_{+}}.  \label{WGB}
\end{equation}%
Using Eq. (\ref{KK}), the expression (\ref{SGB}) on the screen is%
\begin{equation}
S=\left[ \frac{1}{4}A+3\alpha \frac{1-c}{r^{2}}A\right] _{r=r_{+}}.
\label{SGB1}
\end{equation}%
Using Eqs. (\ref{T}), (\ref{WGB}) and (\ref{SGB1}) we have%
\begin{equation*}
TdS+WdV=\frac{3\pi r_{+}}{4}(1-c)dr_{+}.
\end{equation*}%
The generalized Misner-Sharp energy (\ref{EGB}) on the screen is%
\begin{equation}
E=\frac{3\pi r^{2}}{8}\left[ (1-c)+\tilde{\alpha}r^{-2}(1-c)^{2}\right]
_{r=r_{+}}  \label{EGB1}
\end{equation}%
and its variation is%
\begin{equation}
dE=\frac{3\pi r_{+}}{4}(1-c)dr_{+}\text{.}  \label{dEGB}
\end{equation}%
One can find that Eq. (\ref{PGib}) holds on holographic screens actually.
Moreover, this equation will reduce to the first law obtained in \cite{Tian}
for static spacetimes.

\subsection{Hayward's identity of $f(R)$ gravity}

Now we will study whether or not there is the Gibbs equation (\ref{HGib})
for $f(R)$ gravity on general screens in an FRW spacetime. In an FRW
spacetime, the trapping horizon with $\theta _{+}=0$ is located at $\bar{r}=%
\frac{1}{\dot{a}}$, but we will consider a general screen at%
\begin{equation}
\bar{r}=\frac{\sqrt{1-c}}{\dot{a}}\;\text{i.e. }r_{+}=\frac{\sqrt{1-c}}{H}.
\label{roc}
\end{equation}%
The surface gravity on the screen is%
\begin{equation}
\kappa =-e^{\phi }\partial _{u}\partial _{v}r=-\frac{r}{2}(2H^{2}+\dot{H})=-%
\sqrt{1-c}H(1+\frac{\dot{H}}{2H^{2}}).  \label{Kapa1}
\end{equation}

The tangent vector $\xi ^{b}$ (here index $b=t,\bar{r}$) (\ref{zeta}) can be
read as%
\begin{equation}
\xi ^{b}=(1,-\frac{\bar{r}(H^{2}+\dot{H})}{H}),  \label{zeta1}
\end{equation}%
up to a proportional factor. Using the field equation of $f(R)$ gravity (\ref%
{GfR}) and the tangent vector (\ref{zeta1}), the energy supply along the
screen can be obtained as%
\begin{equation}
A\Psi _{b}\xi ^{b}=\left( 1-c\right) ^{\frac{3}{2}}\left[ \frac{f_{R}\dot{H}%
(2H^{2}+\dot{H})}{2H^{4}}-\frac{\dot{f}_{R}(2H^{2}+\dot{H})}{4H^{3}}+\frac{%
\ddot{f}_{R}(2H^{2}+\dot{H})}{4H^{4}}\right] .  \label{Apsi2}
\end{equation}%
The Eq. (\ref{SfR}) in the FRW spacetime can be expressed as%
\begin{equation}
S=\frac{A}{4}(f_{R}-\frac{2H\dot{f}_{R}}{2H^{2}+\dot{H}}).  \label{Sfr}
\end{equation}%
Multiplying the factor $T=\kappa /\left( 2\pi \right) $ to its variation
along the screen, we obtain%
\begin{equation}
T\xi ^{a}\nabla _{a}S=\left( 1-c\right) ^{\frac{3}{2}}\left[ \frac{f_{R}\dot{%
H}(2H^{2}+\dot{H})}{2H^{4}}-\frac{\dot{f}_{R}(4H^{4}+16H^{2}\dot{H}+3\dot{H}%
^{2}+2H\ddot{H})}{4H^{3}(2H^{2}+\dot{H})}+\frac{\ddot{f}_{R}}{2H^{2}}\right]
.  \label{dsfR}
\end{equation}%
It is interesting to find that Eq. (\ref{Apsi2}) is same as Eq. (\ref{dsfR}%
), provided that the radius of screen (\ref{roc}) $r_{+}$ is varied so
slowly that%
\begin{equation}
\dot{H}\ll H^{2},\;\ddot{H}\ll H^{3}.  \label{condition}
\end{equation}%
So we have established the Gibbs equation (\ref{HGib}) for $f(R)$ gravity on
holographic screens. In the general, however, Eq. (\ref{HGib}) does not
hold. The difference between Eq. (\ref{Apsi2}) and Eq. (\ref{dsfR}) can be
given as%
\begin{eqnarray*}
Td_{H}S &\equiv &\xi ^{a}\nabla _{a}E-W\xi ^{a}\nabla _{a}V-\frac{\kappa }{%
2\pi }\xi ^{a}\nabla _{a}S \\
&=&A\Psi _{b}\xi ^{b}-\frac{\kappa }{2\pi }\xi ^{a}\nabla _{a}S \\
&=&\left( 1-c\right) ^{\frac{3}{2}}\left[ \frac{\dot{H}\ddot{f}_{R}}{4H^{4}}+%
\frac{\dot{f}_{R}(6H^{2}\dot{H}+\dot{H}^{2}+H\ddot{H})}{4H^{3}(2H^{2}+\dot{H}%
)}\right] .
\end{eqnarray*}%
It is not clear whether this difference should be interpreted as the
nonequilibrium entropy production invoked in \cite{Eling,Akbar,Cao,Cao3}.

\subsection{Padmanabhan's identity of $f(R)$ gravity}

Next we will check Padmanabhan's identity (\ref{PGib}). In an FRW spacetime,
the work density can be written as%
\begin{equation}
W=\frac{1}{16\pi }\left( f-6f_{R}H^{2}+5H\dot{f}_{R}-4f_{R}\dot{H}+\ddot{f}%
_{R}\right) .  \label{W1}
\end{equation}%
The key step is to consider the variations. Considering $E$ and $S$ on
screens with (\ref{roc}), we notice that the differential $d$ can be
expressed as%
\begin{equation*}
d=dr_{+}\frac{d}{dr_{+}}=dr_{+}\frac{dt}{d\frac{\sqrt{1-c}}{H(t)}}\frac{d}{dt%
}.
\end{equation*}%
Thus, we can replace the differential $d$ with $\partial _{t}$ for the aim
of checking Padmanabhan's identity (\ref{PGib}). Using Eqs. (\ref{GfR}), (%
\ref{SfR}), (\ref{Kapa1}), and (\ref{W1}), one can obtain the right hand of
Eq. (\ref{PGib}) as%
\begin{eqnarray}
TdS+WdV &\sim &T\partial _{t}S+W\partial _{t}V  \notag \\
&=&\left( 1-c\right) ^{\frac{3}{2}}[-\frac{f\dot{H}}{4H^{4}}+\frac{f_{R}\dot{%
H}\left( 5H^{2}+3\dot{H}\right) }{2H^{2}}  \notag \\
&&-\frac{\dot{f}_{R}\left( 2H^{4}+13H^{2}\dot{H}+4\dot{H}^{2}+H\ddot{H}%
\right) }{2H^{3}\left( H^{2}+2\dot{H}\right) }+\frac{\ddot{f}_{R}\left(
2H^{2}-\dot{H}\right) }{4H^{4}}]  \label{left}
\end{eqnarray}%
The left hand is%
\begin{equation}
dE\sim \partial _{t}E=\left( 1-c\right) ^{\frac{3}{2}}\left[ -\frac{f\dot{H}%
}{4H^{4}}+\frac{f_{R}\dot{H}\left( 5H^{2}+3\dot{H}\right) }{2H^{2}}-\frac{%
\dot{f}_{R}\left( H^{2}+3\dot{H}\right) }{2H^{3}}+\frac{\ddot{f}_{R}}{2H^{2}}%
\right] .  \label{right}
\end{equation}%
Comparing Eqs. (\ref{left}) and (\ref{right}) under the approximation (\ref%
{condition}), we have justified the Gibbs equation (\ref{PGib}). In the case
without the approximation, we have%
\begin{eqnarray*}
Td_{P}S &\equiv &dE-\frac{\kappa }{2\pi }dS-WdV \\
&\sim &\partial _{t}E-\frac{\kappa }{2\pi }\partial _{t}S-W\partial _{t}V \\
&=&\left( 1-c\right) ^{\frac{3}{2}}\left[ \frac{\dot{H}\ddot{f}_{R}}{4H^{4}}+%
\frac{\dot{f}_{R}(6H^{2}\dot{H}+\dot{H}^{2}+H\ddot{H})}{4H^{3}(2H^{2}+\dot{H}%
)}\right] .
\end{eqnarray*}%
Interestingly, one can find $d_{P}S\sim d_{H}S$, which suggests both of them
have the same origin and Padmanabhan's approach is consistent with Hayward's
one even off the horizon.

We will further check Padmanabhan's identity (\ref{PGib}) in the static
spacetime where the generalized Misner-Sharp energy is also found. Now we
will evaluate the right hand in Eq. (\ref{PGib}). The surface gravity is $%
\kappa =g^{\prime }/2$ and the Noether charge (\ref{SfR}) is $S=f_{R}A/4$.
Using the field equation (\ref{GfR}), we obtain
\begin{equation}
TdS+WdV=\left( \frac{1}{4}r_{+}^{2}f+r_{+}f_{R}g^{\prime }+\frac{1}{4}%
r_{+}^{2}f_{R}g^{\prime \prime }\right) dr_{+},  \label{right1}
\end{equation}%
where we have considered that $R$, $f$, and $f_{R}$ are all constant, which
is the requirement of quasi-local Misner-Sharp energy. Reading the
generalized Misner-Sharp energy (\ref{EfR}) on the screen and respecting
that $R$, $f$, and $f_{R}$ are all constant, we can get the energy variation
as%
\begin{equation*}
dE=\left( \frac{1}{4}r_{+}^{2}f+\frac{1}{2}f_{R}-\frac{1}{4}%
r_{+}^{2}f_{R}R\right) dr_{+}.
\end{equation*}%
We note $g=c$ is constant in the variation. Substituting Ricci scalar%
\begin{equation*}
R=\frac{2}{r^{2}}-\frac{2g}{r^{2}}-\frac{4g^{\prime }}{r}-g^{\prime \prime },
\end{equation*}%
the variation $dE$ can be recast as%
\begin{equation*}
dE=\left( \frac{1}{4}r_{+}^{2}f+r_{+}f_{R}g^{\prime }+\frac{1}{4}%
r_{+}^{2}f_{R}g^{\prime \prime }\right) dr_{+},
\end{equation*}%
which is same as Eq. (\ref{right1}). Thus, we have shown that Padmanabhan's
identity (\ref{PGib}) holds.

\subsection{Two identities of scalar-tensor gravity}

In the following we will show that the two identities can also be built up
for scalar-tensor gravity on slowly varying screens in an FRW spacetime. We
will use an indirect but more simple procedure compared to the one given in $%
f(R)$ gravity. One may notice that the construction of two identities of $%
f(R)$ gravity on any screens of FRW spacetime are same as the ones on the
trapping horizon up to a factor $\left( 1-c\right) ^{\frac{3}{2}}$. This
implies that the thermodynamic parameters $T$, $W$ and the variation of $U$,
$S$ in two identities should be independent with $c$ up to a factor and the
total factor of each term in two identities should be same. We will show
below that these facts also hold for scalar-tensor gravity. Since we have
obtained the two identities on the slowly varying horizon in an FRW
spacetime \cite{Wu10}, we will finally have both identities of scalar-tensor
gravity on general screens (with slowly varying radius).

The parameter $S$ on general screens in an FRW spacetime has the same form
as Eq. (\ref{Sfr}) replacing $f_{R}$ with $F$. It has a factor $1-c$. The
parameters $U$ (\ref{EST}) and $V$ are dependent with a factor $\left(
1-c\right) ^{\frac{3}{2}}$. The parameter $T$ can be given by Eq. (\ref%
{Kapa1}), which is proportional to a factor $\sqrt{1-c}$. The parameter $W$
can be evaluated as%
\begin{equation*}
W=\frac{1}{16\pi }\left( -V+5H\dot{F}+6FH^{2}+2F\dot{H}+\ddot{F}\right)
\end{equation*}%
which is independent with $c$. So we have found that each terms in two
identities have a total factor $\left( 1-c\right) ^{\frac{3}{2}}$ and the
two identities hold on any screens of FRW spacetime. Note that this is an
another evidence showing the similarity between the $f(R)$ gravity and
scalar-tensor gravity in gravitational thermodynamics.

\section{Second law of modified gravity theories on dynamical holographic
screens}

Hayward showed that for Einstein gravity there is a second law of horizon
entropy \cite{Hayward1}. We review the proof briefly. Denote the tangent
vector to the horizon by $\xi =d/d\lambda =\beta \partial _{u}-\alpha
\partial _{v}$ with norm $2e^{-\phi }\alpha \beta $. Fix the orientations by
$\theta _{+}=0$ and $\beta >0$ on the horizon. Consider $0=d\theta
_{+}/d\lambda =\beta \partial _{u}\theta _{+}-\alpha \partial _{v}\theta
_{+} $, which yields%
\begin{equation*}
dr/d\lambda =-\alpha \partial _{v}r=-\beta r\theta _{-}\partial _{u}\theta
_{+}/2\partial _{v}\theta _{+}.
\end{equation*}%
Using the null energy condition%
\begin{equation}
T_{uu}\geq 0\text{ and }T_{vv}\geq 0\text{,}  \label{NEC}
\end{equation}%
and the Einstein field equation, one can know $\partial _{u}\theta _{+}\leq
0 $. The signs of $\theta _{-}$ and $\partial _{v}\theta _{+}$ are given by
the definition of future or past, outer or inner trapping horizons. Then one
can obtain $\xi ^{a}\nabla _{a}S=2\pi rdr/d\lambda \geq 0$ for future outer
or past inner trapping horizons. Since $0\leq dr/d\lambda =-\alpha \partial
_{v}r$, one can find $\alpha \geq 0$ ($\leq 0$) for future (past) horizons.
Thus, the second law holds as the future outer (respectively past inner)
trapping horizon is perturbed along the spatial (respectively timelike) or
null tangent vector.

Moreover, Hayward also proved the monotonicity of Misner-Sharp energy on
untrapped spheres \cite{Hayward1}, which states that if the dominant energy
condition holds on an untrapped sphere, the Misner-Sharp energy is
non-decreasing in any outgoing spatial or null direction. Since the
Misner-Sharp energy on the horizon for Einstein gravity is $r/2$, Hayward combined the
second law and monotonicity, summarizing that Misner-Sharp energy is
non-decreasing in outgoing directions, defined for untrapped or marginal
surfaces.

In Ref. \cite{Wu10}, we have generalized the second law of horizon entropy
to the modified gravity theories. Here we will further give a similar second
law for the Noether charge (\ref{WaldS}) on the untrapped screen. Our
derivation is motivated by the proof of the monotonicity of Misner-Sharp
energy on untrapped spheres. Moreover, we need to invoke Hayward's identity (%
\ref{HGib}) obtained in the previous section.

Let us fix the orientation by $\theta _{+}>0$ and $\theta _{-}<0$ on the
untrapped sphere. On dynamical holographic screens, consider a spatial or
null tangent vector denoted as $\xi =d/d\lambda =\beta \partial _{u}-\alpha
\partial _{v}$ (Note $\alpha =0$ or $\beta =0$ for null vector.). Since we
have constructed Hayward's identity (\ref{HGib}), one has%
\begin{equation*}
\xi ^{a}\nabla _{a}S=\frac{A\Psi _{a}\xi ^{a}}{T}=\frac{A}{T}\left( \frac{%
-cT_{vv}\alpha }{2r_{,v}}-e^{\phi }T_{uu}\beta r_{,v}\right) =\frac{2\pi A}{%
r_{,uv}}\left( \frac{ce^{-\phi }T_{vv}\alpha }{2r_{,v}}+T_{uu}\beta
r_{,v}\right) .
\end{equation*}%
Consider the null energy condition (\ref{NEC}), the untrapped condition $c>0$%
, the outgoing (ingoing) vector with $\xi ^{a}\nabla _{a}r=-\alpha
r_{,v}+\beta r_{,u}>0$ ($\xi ^{a}\nabla _{a}r<0$), and the outer (inner)
sphere with $r_{,uv}<0$ ($r_{,uv}>0$). One can conclude that the variation
of $S$ is not decreasing along the spatial or null outgoing (ingoing) vector
for outer (inner) untrapped sphere. Thus our dynamical Noether charge (\ref%
{WaldS}) satisfies a second law in the cases where Hayward's identity (\ref%
{HGib}) holds. Similar to the monotonicity of Misner-Sharp energy on
untrapped spheres \cite{Hayward1}, this second law has the physical
interpretation that the entropy contained in an outer (inner) untrapped
screen is non-decreasing as the screen is perturbed outwards (inwards) along
the spatial or null tangent vector of the screen.

\section{Conclusion and discussion}

It can be convinced that there is a deep connection between gravity and
thermodynamics. But since Bekenstein discovered their analogy, the
gravitational thermodynamics is mainly restricted on the horizon. Until
recently, Verlinde proposed that one may associate thermodynamics parameters
on holographic screens, it seems that it is urgent to investigate what is
the role of horizon taken in gravitational thermodynamics or what is the
thermodynamic object of gravity.

In this paper, we investigate the gravitational thermodynamics on dynamical
holographic screens. We are interested on the screen with sphere symmetry,
where the Kodama vector replaces the Killing vector as a preferred direction
of time. One can find that the holographic screen corresponds to the
untrapped sphere.

We discuss the thermodynamics for some typical modified gravity theories.
For Gauss-Bonnet gravity, we have constructed Hayward's identity and
Padmanabhan's identity on any holographic screens. For $f(R)$ gravity and
scalar-tensor gravity, these two indentiteis are built up in the FRW
spacetime with slowly varying screen. For $f(R)$ gravity, we also have
obtained Padmanabhan's identity in the static spherically symmetric
spacetime with constant scalar curvature. We are restricted in these cases
because there exists the generalized Misner-Sharp energy. Since these
identities have the same form as the first law on trapping horizons, we call
them as the first law on holographic screens. Moreover, invoking the
constructed Hayward's identity, we have presented that there is a second law
which holds under the null energy condition for the entropy on the screen.

We argure that the results obtained in this paper support to take general
holographic screens as the thermodynamic objects of gravity. But final
conviction needs further study like Hawking radiation and quantum
statistical model for Noether charge (\ref{WaldS}) on the screens.

\begin{acknowledgments}
SFW, XHG, and PMZ were partially supported by NSFC under Grant Nos.
10905037, 10947116, and 10604024, respectively. XHG was partially supported
by Shanghai Rising-Star Program No.10QA1402300. PMZ was partially supported
by the CAS Knowledge Innovation Project No. KJcx.syw.N2. GHY, SFW and XHG
were also partially supported by Shanghai Leading Academic Discipline
Project No. S30105 and the Shanghai Research Foundation No. 07dz22020.
\end{acknowledgments}

\end{document}